\documentstyle[11pt]{article}

\title{\bf Fusion of strings vs. percolation and the transition to the
quark gluon plasma}
\author{M.A.Braun
\thanks{ On leave of absence from the  Department of High
Energy Physics, University of St. Petersburg, 198904 St. Petersburg, Russia.}
\hspace{3mm} C.Pajares
 \\ Departamento
de F\'{\i}sica de Part\'{\i}culas,\\ Universidade de Santiago de
Compostela,\\ 15706-Santiago de Compostela, Spain\\
and J.Ranft\\
INFN Laboratorio Nazionale del Gran Sasso}
\date{July 1997}
\def\beq{\begin{equation}}
\def\eeq{\end{equation}}
\def\noi{\noindent}
\def\avnu{\langle \nu_{n} \rangle}
\def\avm{\langle M \rangle}
\def\avmu {\langle \mu \rangle}

\begin{document}
\maketitle
\medskip
\noi{\bf Abstract.}

In most of the models of hadronic collisions the number of exchanged
colour strings grows with energy and atomic numbers of the projectile and
target. At high string densities  interaction between them becomes
important, which should melt them into the quark-gluon plasma state in the
end. It is shown that under certain reasonable assumptions about the
the  string interaction, a phase transition to the quark
gluon plasma indeed takes place in the system of many colour strings. It
may be  of the first or second order, depending on the particular mechanism
of the interaction. The critical string density is about unity in both
cases. In the latter case the percolation of strings occurs above the
critical density. The critical density may have been already reached in
central Pb-Pb collisions at 158 A GeV.

\vspace{3.0cm}
%\noi{\Large\bf hep-ph/9707363}\\
\noi{\Large\bf US-FT/24-97}
\newpage
\section{Introduction}

Multiparticle production
at high energies is currently (and successfully) described in terms of
colour strings stretched between the projectile and target [ 1-3].
Hadronizing these strings produce the observed hadrons. In the original
version the strings live independently and the observed spectra are just
the sum of individual string spectra. However, with growing energy and/or
atomic number of colliding particles, the number of strings grows and one
expects that interaction between them becomes essential. This problem
acquires still more importance, considering the idea that at high enough
energy  collisions of heavy enough nuclei may produce the quark-gluon
plasma (QGP). The interaction between strings then has to make the system
evolve towards the QGP state.

In our earlier publications we proposed a simple probabilistic model of
string fusion [4], which was later realized as a Monte-Carlo algorithm
[5]. It turned out to be quite successful in explaining some peculiar
features of  multiparticle production, such as heavy flavour enhancement
at high energies and production of particles outside the kinematical limits
of nucleon-nucleon collisions [6,7]. It did not however seem to predict any
phase transition in the colour string system, which could be interpreted
as a transition to the QGP.

In a recent publication it was noted that the string density reached in Pb-Pb
collisions at present energies is of the same order of magnitude as the
critical density for the percolation of overlapping discs [8]. This could
mean that strings may evolve to the QGP via the percolation process. It
however remained not clear how the percolation can be incorporated into a
more general picture of string interaction (and fusion), and why simple
models of this interaction (in particular, the one proposed in [ 4]) do
not lead to a phase transition. Also one would  like to know possible
alternatives to the percolation to be able to study from the experimental
data  the particular way the hadronic matter chooses to go into the QGP
state.
This is also related to the possible signatures of the percolation or of
any other phase transition.

This note intends to discuss these poroblems. Essentially our conclusions
are as follows. We show that the model of Ref. [4] indeed lacks any phase
transition. Its absence is due to a particular fusion mechanism chosen (Sec.
2). A slight change in this mechanism gives  a very similar and also 
explicitly solvable model, in which there is a clear phase transition to
the QGP. This transition is of the 1st order (Sec. 3). The percolation
occupies an intermediate place between the two models in what concerns the
interaction mechanism. Its predictions for observable quantities are
discussed in Sec. 4. Sec. 5 contains some conclusions.

This note is centered on the analytical treatment of the problem of string
interaction. Such questions as string formation and their hadronization,
which require Monte Carlo simulations, remained outside its scope. As a
consequence we shall study only very primitive observables, whose behaviour
under string fusion can be predicted on plausible assumptions: the total
(charged) multiplicity and the total number of strings. The latter shows
itself in forward-backward correlations [9].
We shall study the behaviour of the system of a fixed number $N$ of strings
under their interactions. In practice, of course, the number $N$
fluctuates, its average value growing with energy. This will undoubtedly
smooth  all irregularities introduced by the phase transition and make its
detection still more difficult.

\section{Fusion of strings in the model [4]}

In this section we briefly review the model proposed in [4] from a
somewhat novel point of view, looking for a possible sign of a phase
transition in it. It was assumed in the model that strings fuse as soon as
their positions in the transverse space coincide. The probability of fusion
$p$ was assumed to be the ratio of the string own transverse area $v$ to
the total interaction transverse area $V$. Of course one should consider
$v$ as an effective area, which may be smaller than the geometrical one
for small fusion probability. We also assumed that colours of fusing strings
(i.e. colours placed at their ends) sum into the total colour of the
resulting string. For the realistic QCD this summation should be performed
according to the composition laws of the $SU(3)$ colour group. This can only
be done within a Monte Carlo code. In the analytical approach of [4], as
well as in the present note, to see qualitative predictions of string
interaction, we adopt a simplified, Abelian, approach, taking that two
strings with positive integer colours $n_1$ and $n_2$ fuse into a string
of colour $n_1+n_2$. Ordinary string possess a unit colour $n=1$. The
basis of the model is laid by the probability $P(\nu_{n})$ to find $\nu_n$
strings of colour $n$, chosen to be
\beq
P(\nu_n)=c\frac{Q^N p^{N-M}}{\prod_{n=1}(\nu_{n}!(n!)^{\nu_n})}
\prod_{k=1}^{M-1}(1-kp)
\eeq
Here $N=\sum_{n}n\nu_n$ is the fixed total colour of the strings which
is also the  number of ordinary strings originally produced; $M=\sum_n
\nu_n$ is the total number of strings ("clusters", in terms of the
statistical approach). In the following we pretend to change this
probability. So it is instructive to analyse its foundation.

Since we shall study the behaviour of the system at fixed $N$, factors
depending only on $N$ do not play any role and are actually a part of the
normalization constant. The denominator factors in (1) are of the standard
quantum statistical origin: $\nu_n!$ takes into account the identity of
$\nu_n$ strings of colour $n$ and each $n!$ takes into account the identity
of $n$ ordinary strings which fuse into a string of colour $n$. The number
$N-M$ is, in fact, the number of fusions, hence the factor $p^{N-M}$. We
come finally to the factor
\beq w=\prod_{k=1}^{M-1}(1-kp)\eeq
which is the one that contains information about a particular way strings
fuse.

To obtain it we made an additional assumption that fusion does not change
the proper volume $v$ of strings, that is, all strings have the same volume
$v$ irrespective of their colour (and consequently the "string tension"
of a string with colour $n$ is $n^2$ greater than of the ordinary string).
As a result, fused strings have the same probability to fuse as ordinary
ones. The factor $w$ can then be obtained from a simple recurrency
relation, which follows if one studies the system with one string of
arbitrary color $k$ added. For this system to exist, it is necessary that
the new string should not fuse with the rest. This leads to a relation
\beq
w(\nu_n+\delta_{nk})=w(\nu_n)(V-v\sum_n \nu_n )/V
\eeq
The second factor, equal to $1-p\sum_n \nu_n$, gives precisely the
probability not to fuse with the rest. We chose to write it in terms of
volumes to compare with the treatment in the following section. The
recurrency relation (3) has evidently the solution (2)

Given the probability (1), one can calculate all observables of interest.
Our main observable will be the average number of strings of colour $n$,
$\avnu$. One easily finds (see Appendix for some details)
\beq
\avnu\equiv \sum_{\{\nu_n\}}\nu_n
P(\nu_n )=C_{N}^{n}p^{n-1}(1-p)^{N-n}
\eeq
It is just a binomial distribution in $n$. With the help of (4) we obtain
the average total number of strings $\avm$:
\beq
\avm=\frac{1}{p} \left(1-(1-p)^{N}\right)
\eeq
From this last equation we see that $\avm$ monotonously falls from $\avm=N$
at $p=0$ to $\avm=1$ at $p=1$, which clearly shows the effect of the
fusion process but exhibits no sign of irregularity, which could indicate a
phase transition.

One would also like to know the behaviour of the average multiplicity
$\avmu$ of particles produced by the hadronizing strings as a function of
$p$. To this purpose one has to make certain assumptions about the
multiplicity $\mu_n$ corresponding to a string of colour $n$. On physical
grounds the multiplicity should be proportional to the surface of the string
area. With all strings having the same dimension this does not lead to any
$n$ dependence. One may also expect some growth of the multiplicity 
with $n$ from the
growing string tension. In the limiting case when the string of colour $n$
emits particles exactly as $n$ ordinary strings we shall have
\[ \avmu=\nu_1\sum_n n\nu_n =N\mu_1\]
that is, one will not see any change in the spectra resulting from fusion.
This is not a surprise, since with $\mu_n=n\mu_1$ fusion does not influence
the multiplicity. In the opposite case, when a high-colour string emits
particles exactly like an ordinary one, $\mu_n=\mu_1$ and one has
\[\avmu=\mu_1 M\]
and with the growth of $p$ $\avmu$ falls from $N\mu_1$ to $\mu_1$.

More generally, one can study the distribution in the number $m$ of
produced particles, assuming that this distribution is $\lambda_n (m)$ for
a string of colour $n$, satisfying
\[\sum_m \lambda_n(m)=1;\ \ \sum_m m\lambda_n(m)=\mu_n\]
The resulting distribution $\Lambda(m)$ is then given by the average
\beq
\Lambda(m)=\frac{\sum_n\lambda_n(m)<\nu_n>}{\sum_n <\nu_n>}
\eeq
From (6) one can calculate moments of the distribution $\Lambda(m)$. In
particular the average number of emitted particles per string is
\beq
<m>=<\mu>/<M>
\eeq
and the dispersion squared is
\beq
D^2=\sum_m m^2\Lambda(m)-<m>^2
\eeq
Presenting 
\beq
\sum_m m^2\lambda_n(m)=\mu_n^2+d_n^2
\eeq
where $d_n$ is evidently the dispersion of the distribution in $m$ for a
string of colour $n$, we split $D^2$ into an internal part, generated by
this intrinsic dispersion, and an external one, generated by fluctuations
in the distribution of strings in colour. This latter is given by
\beq
D^2_{ex}=\sum_n\mu_n^2<\nu_n>-<m>^2
\eeq and is presumably sensitive to a possible phase transition. It is
positive and vanishes both for $p=0$ and $p=1$.  

To have a clearer vision of the model from the statistical point of view,
it is instructive to consider the "thermodinamical limit", which in our
case means $V\rightarrow\infty$ and $N\rightarrow\infty$ with the density
of strings $\rho=N/V$ constant. Evidently then $p=v/V\rightarrow 0$ and the
relevant parameter describing string interaction is
\beq
\eta=\rho v=Np
\eeq
In this limit we find from (4) and (5)
\beq
\avnu/N=\frac{1}{\eta\sqrt{2\pi n}}\exp
\left(n-\eta+n\ln\frac{\eta}{n}\right)
\eeq
and
\beq
\avm/N=\frac{1}{\eta}\left(1-e^{\eta}\right)
\eeq
We observe that the model has  correct scaling properties with $N$: both
$\avnu/N$ and $\avm/N$ result independent of $N$ in the limit. From (12) we
conclude that $\avnu$ has a maximum at $n=\eta$, so that with the growth of
$\eta$ the distribution steadily shifts towards higher $n$. However there
is no phase transition: the system does not change qualitatively as
$\eta\rightarrow\infty$.

 This circumstance is clearly seen from Figs. 1 to
3, which show $<M>$, $<\mu>$ (in units $\mu_1$) and $D^2_{ex}$ (multiplied
by 10) as a function of the parameter $\eta$ in the system of 50 ordinary
strings. In this figures prediction of the model studied in this section are
illustrated by the upper curves (marked 1). The curves are quite smooth and
show a very slow fall off of  $<M>$ and
$<\mu>$ and an equally slow growth of $D^2_{ex}$ with the growth of $\eta$.

The reason of this is, of course, evident. It resides in the assumption
that the dimension of strings does not grow with colour. Strings with
higher colour then fuse with the same probability as ordinary ones, so
that fusion is not accelerated with the accumulation of high colour
strings. This picture does not correspond to what one might expect on 
intuitive grounds. Presently we shall see, that once we substitute this
assumption by a more physical one, the behaviour of the system changes
radically and acquires the expected features with a clear phase transition
at high enough $p$.

\section{Fusion of strings with the conservation of their volume: 1st
order phase transition}

An alternative model, which also preserves the important quality of being
explicitly soluble, is obtained if we assume that as string fuse, they
conserve their proper volume. Namely, two ordinary strings fuse into a
string with the volume $2v$. Then generally a string of colour $n$ will
have volume $nv$, and consequently the probability to fuse much higher.
To obtain the relevant factor $w$ we construct a recurrency relation
similar to (3). However in this case the factor $w$ will depend not only
on the occupation numbers $\nu_n$ but also on the volume $V$, since the
probability of fusion depends on the volume available and the latter now
depends on the colour of the added string. The recurrency relation we
obtain is
\beq
w(V,\nu_n+\delta_{nk})=c\,w(V-kv,\nu_n)(V-kv-v\sum_n \nu_n )
\eeq  
We have taken into account that the rest strings live in a smaller volume
$V-kv$ left after a string of colour $k$ has been added.

Recurrency relation (14) can also be solved quite easily. We rewrite it as
\beq
w(V+kv,\nu_n+\delta_{nk})=cw(V,\nu_n)(V-v\sum_n \nu_n )
\eeq  
Note that the factor $V-v\sum_n \nu_n$ is invariant under a simulataneous
change $V\rightarrow V+kv$, $\nu_n\rightarrow \nu_n+\delta_{nk}$.
Therefore, making $q_k$ such substitutions we obtain
\beq
w(V+kq_k v,\nu_n+q_k\delta_{nk})=c^{q_k}w(V,\nu_n)(V-v\sum_n \nu_n )^{q_k}
\eeq
and more generally
\beq
w(V+v\sum_{k}kq_k,\nu_n+q_n)=c^{\sum_{k}q_k}w(V,\nu_n)(V-v\sum_n \nu_n
)^{\sum_{k}q_k}
\eeq
Putting finally $\nu_n =0$, $w(V,0)=1$ and $c=1/V$ we obtain
\beq
w=(1-pN)^{M}
\eeq
Thus the distribution in $\nu_n$ which desribes our new model is
\beq
P(\nu_n)=c\frac{Q^N p^{N-M}}{\prod_{n=1}(\nu_{n}!(n!)^{\nu_n})}
(1-pN)^{M}
\eeq

One observes at once that this distribution makes sense only for $p\leq
1/N$, or, in terms of the parameter $\eta$, Eq. (11)
\beq
\eta\leq 1
\eeq
What happens if $p$ becomes greater than $1/N$? The answer is evident. The
total volume of strings $Nv$ then exceeds the interaction volume $V$ and the
only way for the strings to exist is to form a single string which spans
all the interaction area. In other words, with the
probability one, all string fuse into what may be viewed as the QGP plasma.
This transition is of the first order, as we shall presently see considering
the thermodynamic limit.

With the distribution (19) one can also calculate average occupation numbers
in the (nearly) explicit form. One finds (see Appendix)
\beq
\avnu=N\alpha C_{N}^{n}\frac{f_{N-n}(N\alpha)}{f_{N}(N\alpha)}
\eeq
and 
\beq
\avm =\frac{f_{N+1}(N\alpha)}{f_{N}(N\alpha)}-N\alpha
\eeq
where
\beq
\alpha=1/\eta -1
\eeq
and $f_{N}(\alpha)$ is defined to be
\beq
f_{N}(\alpha)=\left[ (\frac{d}{dx})^{N}e^{\alpha (e^{x}-1)}\right]_{x=0}
\eeq
It is a polinomial of order $N$.

 These exact expressions are not very
transparent. However one clearly sees from (22) that with the growth of
$\eta$ from 0 to 1 (that is $p$ growing from 0 to $1/N$) the average value
$\avm$ diminishes from $N$ down to exactly 1. Indeed, at $\eta\rightarrow
0$ we have $\alpha\rightarrow\infty$ and only two highest powers of the
polinomials $f_N$ have to be taken into account. One finds
\[ f_N(\alpha)\simeq \alpha^{N}+(1/2)N(N-1)\alpha^{N-1}+...\]
Putting this into (22) one obtains $\avm=N$. At $\eta\rightarrow 1$ we have
$\alpha\rightarrow 0$ and only the lowest term of $f_n$ contributes, which
is $\alpha$ independently of $N$. From (22) one then finds $\avm =1$.

To clearly see the phase transition we again take the thermodynamic limit
$V,N\rightarrow\infty$ with the density $\rho$ and parameter $\eta$ fixed.
Applying the saddle point technique (see Appendix) we then find a
simple formula for $\avnu$
\beq
\avnu/N=\alpha u^n /n!
\eeq where $\alpha$ is given by (23) and $u$ is a solution of the
transcendental equation
\beq
\alpha u=e^{-u}
\eeq
Eq.(25) is nothing but a Poisson distribution (correctly normalized because
of (26)). For the average $\avm$ we obtain
\beq
\avm/N=1/u-\alpha
\eeq
As we see, the system, again, possesses normal scaling properties with 
respect to the volume: the right-hand sides of (25) and (27) are independent
of
$N$.

From (25) one concludes that there is no infinite cluster for $\eta$ smaller
than the critical value $\eta_c=1$, since $\nu_{\infty}=0$. At
$\eta=\eta_c$,
$\alpha$ goes to zero and $u$ becomes infinite, so that Eq.(25) ceases to
exist. At $\eta>\eta_c$ we have a single infinite cluster and no finite
ones. Thus if $P_{f(\infty)}$ are the probabilities to find a finite
(infinite) cluster we have
\[ P_f=1,\ \ \ P_{\infty}=0,\ \ \ 0\leq\eta\leq\eta_{c}=1\]
\[ P_f=0,\ \ \ P_{\infty}=1,\ \ \ \eta\geq\eta_{c}=1\]
This situation corresponds to a 1st order phase transition at $\eta=\eta_c
=1$. Slightly below the phase transition point we have
\beq
\avm/N=\frac{1}{|\ln \left(\alpha\ln\frac{1}{\alpha}\right)|},\ \ 
\alpha\simeq 1-\eta\rightarrow 0
\eeq
so that the behaviour near the critical point is logarithmic, and not
power-like, as for the 2nd order phase transitions.

As to the multiplicity, it is reasonable to assume that it is proportional
to the emitting surface and therefore to $\sqrt{n}$ for the string of
colour $n$. Then the observed average multiplicity will be
\beq
\avmu=\mu_1\sum_{n}\sqrt{n}\avnu
\eeq
$\mu_1$ being as before, the multiplicity for a single ordinary string.
The sum (29) cannot be done analytically for finite $N$. For large $N$ the
saddle point method gives
\beq
\avmu/N=\mu_1/\sqrt{u}
\eeq
where $u$ is the solution of (26). This expression is valid for not very
small $u$: evidently for $\eta=0$ and consequently $u=0$ we have
\beq
\avmu/N=\mu_1
\eeq
At $\eta$ close to the critical point we have
\beq
\avmu/N=\frac{\mu_1}{\sqrt{|\ln \left(\alpha\ln\frac{1}{\alpha}\right)|}}
\eeq
logarithmically small, as the number of strings. Above the phase transition
the whole interaction area acts as a single emitter and we have
\beq
\avmu/N=\frac{\mu_1}{\sqrt{N}}
\eeq
so that the multiplicity is down by a (large) factor $\sqrt{N}$.

These features of $<M>$ and $<\mu>$ are clearly seen in Figs. 1 and 2
(curves marked 2). One finds a sharp drop of both $<M>$ and $<\mu>$ at the
phase transition point $\eta=1$. Substantially below the transition point the
fall off of both these quantities is quite slow and for $\eta$ below 0.5
predictions of the model practically coincide with those of the preceding
section. The same conclusions can be drawn from Fig. 3 where the curve
marked 2 shows the dispersion squared $D^2_{ex}$. Its sharp rise 
as $\eta$ approaches unity and subsequent fall to zero at $\eta$ exactly
equal to unity are quite indicative of the first order phase transition.

\section{Percolation}

An alternative way to the formation of the QGP in the colour string system
is the percolation. This is a purely classical mechanism. Consider
projections of ordinary strings onto the transverse space as filled circles
(discs) of area $v$. Let them be distributed uniformly in the interaction
area $V$. Some of the discs may overlap and in this way form clusters of
string matter. At some critical disc density $\rho$ a cluster appears which
spans all the volume (an infinite cluster if the number of discs
$N\rightarrow\infty$). This infinite string cluster may be identified with
the GQP filling part of the total volume. 

Comparing this mechanism with the string fusion we note that it occupies an
intermediate position between the two scenarios discussed in Secs. 2 and
3, as far as the interaction mechanism is concerned. Indeed, in the
percolation, strings are allowed to partially overlap, so that the volume of
the resulting clusters varies between the two extremes: the volume of a
single string, in the complete overlap case, and the sum of the interacting
strings volumes, in the minimal overlap case. The two scenarios discussed
above  correspond just to these two extreme cases. An additional advantage
of the percolation treatment is that it introduces various clusters for a
given high colour $n>1$, thus taking into account (in a classical, not
quantum way) possible excited states of fused strings, which should become
numerous for large $n$. Unfortunately, the percolation cannot be treated
analytically and nearly all the  results known in the literature have been
obtained by Monte Carlo simulations in the vicinity of the phase transition
point.

The percolation phase transition is known to be of the 2nd order. The
critical value of the parameter $\eta$ introduced above is found to be [10]
\beq
\eta_c\simeq 1.12
\eeq
This value is remarkably close to the critical value $\eta_c=1$ for the 1st
order phase transition in our model of string fusion  discussed in Sec. 3. 
It is also known that the fraction  $\phi$ of the volume, occupied by
strings is determined by the formula [11]
\beq
\phi=1-e^{-\eta}
\eeq
which coincides with the fraction of the volume $\phi=v\avm$ in our string
fusion model of Sec. 2 (Eq. (13)). In the model of Sec. 3, of course, $\phi
=1$, since the volume is conserved in this model and, in terms of
percolating discs, no overlapping takes place.

Below the phase transition point, for $\eta<\eta_c$, there is no infinite
cluster, so that
\beq
P_f=(1/N)\sum_n n\nu_n=1,\ \ \ P_{\infty}=0
\eeq
Above the transition point, at $\eta>\eta_c$ an infinite cluster appears
with a probability
\beq
P_{\infty}=\theta (\eta-\eta_c) (\eta-\eta_c)^{\beta}
\eeq
The critical exponent $\beta$ can be calculated from Monte Carlo
simulations.
However the universality of critical behaviour, that is its independence of
the percolating substrate, allows to borrow its value from lattice
percolation, where [10] 
\[ \beta=5/36\]

The main quantity of interest to us will, as before, be the average
occupation numbers $\avnu$. Their behaviour at all values of $\eta$ and $n$
is not known. From scaling considerations in the vicinity of the phase
transition  it has been found [12]
\beq
\avnu= n^{-\tau}F(n^{\sigma} (\eta-\eta_c)),\ \ |\eta-\eta_c|<<1,\ n>>1
\eeq
where $\tau=187/91$ and $\sigma=36/91$ and the function $F(z)$ is finite
at $z=0$ and falls off exponentially for $|z|\rightarrow\infty$.
Expression (38) is of limited value, since near $\eta=\eta_c$ the bulk of the
contribution is still supplied by low values of $n$, for which (38) is not
valid. However from (38) one can find non-analytic parts of various
observables
at the transition point. In particular, we find the singular part of $\avm$
in the form
\beq
\Delta\avm=c|\eta-\eta_c|^{8/3}
\eeq
This singularity is quite weak: not only $\avm$ itself but also its two
first derivatives in $\eta$ stay continuous at $\eta=\eta_c$ and only the
third blows up as $|\eta-\eta_c|^{-1/3}$.

A little stronger singularity is found in the average multiplicity.
Assuming it again
to be proportional to linear dimensions of strings in the transverse plane,
that is, to $\sqrt{n}$, we get from (38)
\beq
\Delta\avmu=c|\eta-\eta_c|^{101/72}
\eeq
So already the second derivative of $\avmu$ blows up at the transition
point.

More detailed predictions for the total $\avm$ and $\avmu$ , and not only
for their singular parts, can only be obtained from Monte Carlo simulations.
For a system of 50 strings they are presented in Figs. 1 to 3. by the
curves marked 3. One observes that the behaviour of both $<M>$ and $<\mu>$
is quite smooth and does not seem to show any sign of irregularity at the
phase transition point $\eta\simeq 1.12$. This is of no wonder, considering
very weak singularities (39) and (40) introduced by the phase transition.
Nevertheless, on the whole, the behaviour of both quantities is very
different comparing to the previous models. Below the phase transition the
string fusion  is much stronger in the percolation case, so that averages
of both $M$ and $\mu$ result much smaller than in the fusion models of Secs.
2 and 3. Above the phase transition point these averages continue to fall
off quite smoothly, so that they become substantially greater than for the
1st order transition (Sec. 3) but stay much smaller than in the model of
Sec. 2 without a phase transition. This opens up some possibilities to
study the interaction mechanism from the observed behaviour of $<M>$ and
$<\mu>$.  The dispersion curve in Fig. 3 is more informative. It shows a
clear maximum at values of $\eta$ in the interval 1.1-1.5. The absolute
value of the dispersion at the maximum is nearly two orders of magnitude
higher than those for the fusion models of Secs. 2 and 3 (which are shown in
Fig. 3 multiplied by 10). Such a high dispersion is typical for a phase
transition. From Fig. 3 one can approximately determine the
critical value $\eta_c$ to lie around 1.3. It is somewehat higher than the
value 1.12 known in the literature, but the latter corresponds to 
infinitely many strings, and our results refer to only 50. 

\section{Conclusions}
We have shown that string fusion may lead to a phase transition into the
QGP for high enough string density. We presented a particular completely
solvable string fusion model in which this phase transition is of the 1st
order. Percolation is another alternative, which has certain advantages in
treating the details of string interaction. It is important that in both
models the transition to QGP starts practically at the same value of the
string density (corresponding to $\eta=1$ in the string fusion model and
$\eta=1.12$ in the percolation model). So, irrespective of the interaction
mechanisms, one can expect QGP formation at such densities. As pointed out
in [8], this  critical density seems to have been nearly reached in Pb-Pb
collisions already at present energies at SPS and should be considerably
surpassed in future RHIC and LHC  experiments.

We have dealt only with qualitative features of string interactions and
transition to QGP, which allow for an analytical treatment. Detailed
predictions require a Monte Carlo algorithm.

The two models of phase transition presented above seem to be complementary
for such a program. The string fusion model of Sec. 3 is well suited for
the hadronization part, since one can comparatively easily extend the
standard formalism of string decay to strings of higher colour. In fact,
this already has been included in the Monte Carlo string fusion scheme of
Ref [5]. However the model is difficult to realize on the level of string
production: somehow the original strings should be melted into the fused
one of the same geometrical form, which is not at all easy in the
essentially classical Monte Carlo approach. The percolation treatment, on
the contrary, is quite easily and naturally realizable at the string
formation stage, but meets with serious problems at the hadronization
level, when one has to introduce the decay of complicated geometrical
structures corresponding to percolating clusters. These problems are left
for future studies. In any case, experimental signatures like
$J/\Psi$ suppression in Pb-Pb collisions, seem to indicate that percolation
of strings already occurs [8].

\section{Acknowledgements}

This work has been done under the contract AE96-1673 from CICYT (Spain).
M.B. and J.R.thank the University of Santiago de Compostela for 
hospitality and financial support.

\section{References}

\hspace{0.6cm}1.  A. Capella, U. P. Sukhatme, C.--I. Tan and J. Tran Thanh
Van, Phys. Lett. {\bf B81} (1979) 68; Phys. Rep. {\bf 236} (1994) 225.

2.A.B.Kaidalov and K.A. Ter-Martirosyan, Phys. Lett. {\bf B117} (1982) 247.

3.  B. Andersson, G. Gustafson and B. Nilsson--Almqvist, Nucl.
Phys. {\bf B281} (1987) 289.

4. M.A.Braun and C.Pajares, Nucl. Phys. {\bf B390} (1993) 542, 549.

5.  N. S. Amelin, M. A. Braun and C. Pajares, Phys. Lett. {\bf 
B306} (1993) 312;  Z. Phys. {\bf C63} (1994) 507.

6. N.Armesto, M.A.Braun, E.G.Ferreiro and C.Pajares, Phys. Lett. {\bf 344B}
(1995) 301.

7. N.Armesto, M.A.Braun, E.G.Ferreiro, C.Pajares and Yu..M.Shabelski, Phys.
Lett. {\bf 389B} (1996) 78; Astropartcle Phys. {\bf 6} (1997) 329.

8. N.Armesto, M.A.Braun, E.G.Ferreiro and C.Pajares,
Phys. Rev.Lett. {\bf 77} (1996) 3736.

9. N.Amelin, N.Armesto, M.A.Braun, E.G.Ferreiro and C.Pajares, Phys.
Rev. Lett. {\bf 73} (1994) 2813.

10. M.B.Isichenko, Rev. Mod. Phys.{\bf 64} (1992) 961.

11. V.K.S.Shante and S.Kirkpatrick, Adv. Phys. {\bf 20} (1971) 325.

12. D.Stauffer, Phys. Rep. {\bf 54} (1979) 2.
\newpage

\section{Appendix: Averages over the distributions (1) and (14) }.

We illustrate our technique by calculating the normalization constant $c$
in the distibutions (1) and (19).
We use the identity
\beq
\delta_{kl}=\int \frac{dz}{2\pi iz}z^{k-l}
\eeq
where the integration is supposed to go along a closed contour around the
origin. Then for both distributions we obtain
\beq
c^{-1}=(Qp)^{N}\sum_{M}\int \frac{du}{2\pi iu}u^{-N}
\int \frac{dz}{2\pi iz}z^{-M}p^{-M}w(N,M)\sum_{\{\nu_n\}}\frac{
u^{\sum_{n}n\nu_{n}}z^{\sum_{n}\nu_n}}
{\prod_{n=1}(\nu_{n}!(n!)^{\nu_n})}
\eeq
where the sums over $\nu_n$ are now unconstrained and can be easily done.
In fact, we have
\beq
T_n\equiv\sum_{\nu_n}\frac{u^{n\nu_n}z^{\nu_n}}{\nu_n! (n1)^{\nu_n}}=
\exp (zu^{n}/n!)
\eeq
and the product over $n$ gives
\beq
T=\prod_{n=1}T_n=\exp\left(z(e^u-1)\right)
\eeq
Integration over $z$ is trivially performed and gives
\[(e^u-1)^M/M!\]

Now we have to sum over all $M$. This summation is different for the two
distributions (1) and (19). For (1) the factor $w$ is given by (2) and
can be presented in the form
\[ w(M)=p^M M! C_{1/p}^{M}\]
 Then the sum over $M$ gives
\beq e^{u/p}\eeq
For (19) the factor $w$ is given by (18). The summation over $M$ then gives
a more complicated factor
\beq e^{N\alpha(e^u-1)}\eeq
where $\alpha$ is given by (23).

Finally the integration over $u$ has to be done, which reduces to taking
the $N$th term of the expansion of expressions (45) or (46) around
$u=0$. For the distribution (1) it is trivial and gives
\beq
c^{-1}=Q^N/N!
\eeq
whereas for the distribution (19) one obtains
\beq
c^{-1}=(Qp)^N f_N (N\alpha)/N!
\eeq
where $f_N$ is defined by (24).

Note that the polinomials $f_N(\alpha)$ can be easily calculated by a
recurrency relation which relates the coefficients for $f_N$ and $f_{N+1}$.
Indeed one notices that the derivatives of the function (46) have the
structure
\beq
(d/du)^{N}e^{\alpha(e^u-1)}=e^{\alpha(e^u-1)}f_{N}(\alpha e^u)
\eeq
Differentiating once more, one gets a relation 
\beq
c_{N+1}^{(k)}=c_N^{(k-1)}+kc_{N}^{(k)}, 1\leq k\leq N
\eeq
for the coefficients $C_N^{(k)}$ of the polinomial $f_N$. Also one
evidently has
$c_N^{(0)}=0$ and $c_N^{(N)}=1$. Thus, starting from $N=1$, one can
successively calculate all polinomials $f_{N}$ using the recurrency
relation (50).

Averages like $\avnu$ or $\avm $ have the same representation as (42)
 with some simple additional factors and are calculated in a similar manner.

To study the limit $N\rightarrow\infty$ the integral representation for
$f_N(N\alpha)$ is more convenient:
\beq
f_N(\alpha)=\int \frac{du}{2\pi iu}u^{-N}e^{N\alpha(e^u-1)}
\eeq
Presenting the integrand in the form $\exp NE(u)$ we have at large $N$
\[ E\simeq \alpha(e^u-1)-\ln u\]
so that the saddle point is determined by Eq. (26). Knowing the saddle point
it is trivial to calculate the averages $\avnu$, since for them an extra
factor $N\alpha u^{n}/n!$ has to be introduced under the integral. Taking
this factor at the stationary point one obtains Eq. (25).

\newpage
\section{Figure captions}

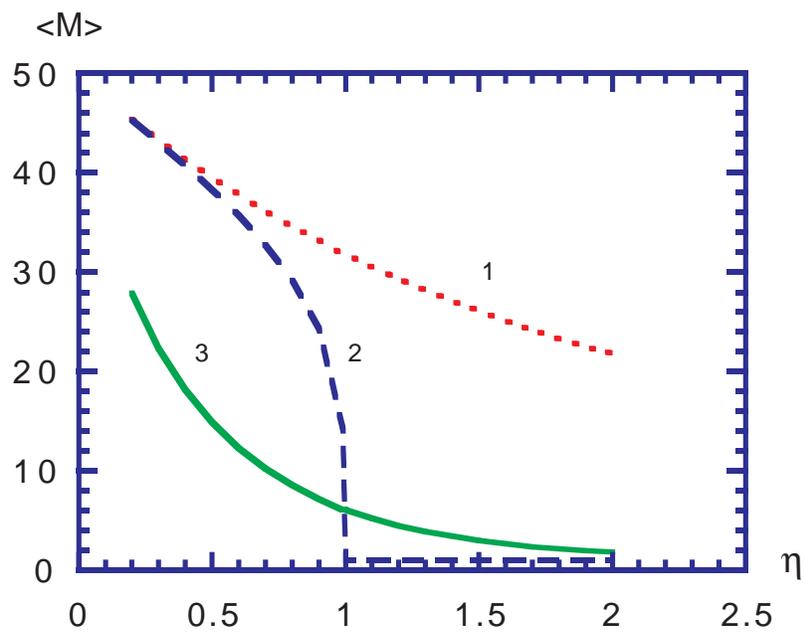
{\bf Fig.1.}  The average  number of clusters $M$ in the system of 50
strings as a function of  $\eta$. Curves 1, 2 and 3 correspond to models of
Secs. 2, 3 and 4 respectively. The latter curve was obtained by Monte Carlo
simulations.\\
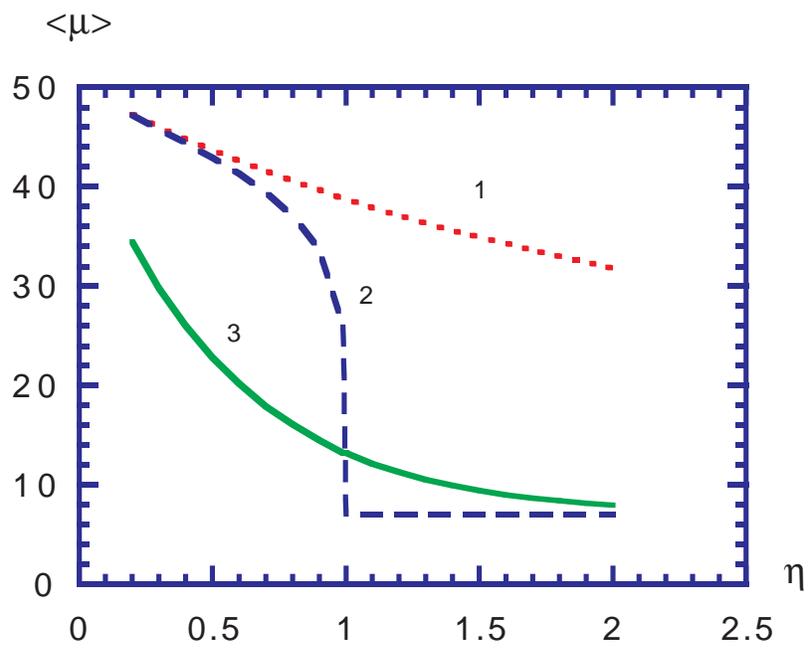
 {\bf Fig.2.}  The average  multiplicity  $\mu$ in units $\mu_1$ in
the system of 50 strings as a function of  $\eta$.  Curves 1, 2 and 3
correspond to models of Secs. 2, 3 and 4 respectively. In all cases $\mu_n$
was assumed proportional to $\sqrt{n}$. The percolation curve (3)  was
obtained by Monte Carlo simulations.\\
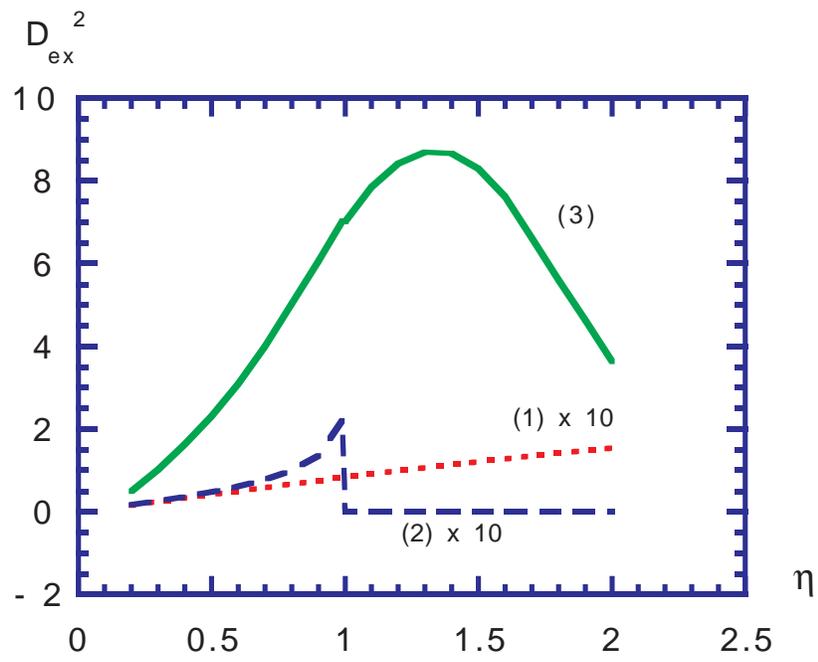
{\bf Fig.3.}  The dispersion squared $D^2_{ex}$ in the system of 50
strings as a function of  $\eta$. Curves 1, 2 and 3 correspond to models of
Secs. 2, 3 and 4 respectively. Curves 1 and 2 show values of $D^2_{ex}$
multiplied by 10.  Curve (3) was obtained from  Monte Carlo simulations
for $<M>$ and $\avmu$ (Figs. 1 and 2).\\

\end{document}